\documentclass[twocolumn]{revtex4}
\usepackage{graphicx}
\begin{document}
\title{The joys of permutation symmetry: direct measurements of entanglement}
\author{S.J. van Enk$^{1,2}$}
\affiliation{
$^1$Department of Physics, University of Oregon\\
Oregon Center for Optics and
Institute for Theoretical Science\\
 Eugene, OR 97403\\
$^2$Institute for Quantum Information, California Institute of Technology, Pasadena, CA 91125\\
}
\begin{abstract}
So-called {\em direct} measurements of entanglement are collective measurements on multiple copies of a (bipartite or multipartite) quantum system that directly provide one a value for some entanglement measure, such as the concurrence for bipartite states. Multiple copies are needed since  the entanglement of a mixed state is not a linear function of the density matrix.  Unfortunately, so far all experimental implementations of direct measurements made unverified assumptions about the form of the states, and, therefore, do not qualify as entanglement verification tests. I discuss how a direct measurement can be turned into a quantitative entanglement verification test  by exploiting a recent theorem by Renner (R.~Renner, Nature Physics {\bf 3}, 645 (2007)).
\end{abstract}
\maketitle
Quantum information theory has produced a handful of different entanglement verification tests [for recent reviews, see \cite{guehne,elk}], among which Bell inequality tests \cite{bell,chsh} and entanglement witnesses \cite{ew} are the best known.  Such measurements are performed on single copies \footnote{When I use the word ``copy'' I do not imply any assumption about the ``copies'' being identical.} of the bipartite system under investigation. One assumes the validity of quantum mechanics, but---and this is a crucial point---nothing about the states to be tested. Of course, in order to obtain sufficient statistics for reliable estimates, a large ensemble of single copies is needed.

Several so-called {\em direct} measurements of entanglement have been considered recently \cite{carteret,horodecki,horodecki2,walborn,harald}. 
Such direct measurements directly measure some entanglement measure \cite{vidal}, such as the concurrence of bipartite states. Since such functions are nonlinear functions of the density matrix, one needs {\em collective} measurements on multiple copies. 

The way these direct measurements are formulated does require an assumption about the form of the state of the multiple copies, namely that one has independent and identical copies. That is, the state of $N$ copies of a multipartite system is assumed to be of the form $\rho_0^{\otimes N}$, for some single-copy multipartite state $\rho_0$. Here I show how a direct measurement of entanglement can be performed such that this assumption on the form of the state of multiple copies is {\em approximately} enforced by taking specific precautions about the way the measurements are performed. The error one makes because of this approximation is analyzed in some detail to show how one obtains a reliable estimate of entanglement from the data gathered. This analysis relies on a recent representation theorem proven by Renner \cite{renner}.

Experimental implementations to date of direct measurements, Refs.\cite{walborn,harald},  relied explicitly on  unverified and unqualified assumptions about the form of the states generated, and thus cannot be considered to be entanglement verification tests. In particular, Ref.~\cite{walborn} made the 
assumption that two copies of the states to be tested were pure and identical. This assumption is rather 
strong: all pure states, except for a set of measure zero, are entangled. All data taken in the experiment are consistent with unentangled states \cite{enkcom,elk}.
 The analysis of entanglement in a more recent experiment \cite{harald} makes the explicit assumption that one has two {\em independent} (although not necessarily identical) copies 1 and 2 of the bipartite system of qubits \footnote{One should also not simply {\em assume} one has qubits, but that is another story.} 
$A$ and $B$, located in Alice's and Bob's labs, respectively. That is, the state of the 4 systems together is assumed to be of
the form
\begin{equation}\label{form}
\rho=\rho_1^{AB}\otimes\rho_2^{AB}.
\end{equation}
But this is a restrictive assumption as it ignores any possible correlations (or entanglement) between the two copies. 
The quantities measured in \cite{harald} are expectation values of two operators
\begin{equation}
V_1=4(P^A_--P^A_+)\otimes P^B_-,
\end{equation}
and
\begin{equation}
V_2=4P^A_-\otimes (P^B_--P^B_+).
\end{equation}
Here, $P^{A}_-$ and $P^B_-$ are the projectors onto the antisymmetric subspaces of the two quantum system 1 and 2 in Alice's and Bob's labs, respectively. Similarly $P^{A}_+$ and $P_+^B$ are the projectors onto the fully symmetric subspaces of the two systems 1 and 2 in Alice's and Bob's labs.
Now if $\rho$ is {\em assumed} to be of the special form (\ref{form}) then one may derive a bound
\begin{equation}\label{bound}
C(\rho_1)C(\rho_2)\geq {\rm Tr}(\rho V_i)
\end{equation}
for $i=1,2$.
But without the assumption (\ref{form}) on $\rho$ the bound is invalid. To see this, consider the following state of four qubits, 
\begin{eqnarray}\label{ce}
|\psi\rangle_{A_1A_2B_1B_2}&=&\frac{1}{2}(|01\rangle_{A1A2}-|10\rangle_{A1A2})\nonumber\\
&&\otimes(|01\rangle_{B1B2}-|10\rangle_{B1B2}).
\end{eqnarray}
Here the first ket refers to the systems in Alice's lab, the second ket refers to the systems in Bob's lab. The state is written such that it is manifestly clear that there is no entanglement between Alice's and Bob's systems.
The observables $V_1$ and $V_2$  have expectation values in the state $|\psi\rangle$ given by
\begin{equation}
\langle\psi| V_i|\psi\rangle=4.
\end{equation}
The bound (\ref{bound}) is violated in the maximum possible way (for states of qubits), with the left-hand side (which is now defined in terms of reduced density matrices $\rho_1={\rm Tr}_2|\psi\rangle\langle\psi|$
and $\rho_2={\rm Tr}_1 |\psi\rangle\langle\psi|$) being zero, and the right-hand side being equal to four.
Thus, a measurement of the observables $V_i$ {\em cannot}  lead to any conclusion about entanglement without assumptions about the state generated. Hence, all data from Ref.~\cite{harald} are consistent with states that have zero entanglement between $A$ and $B$.

The question arises what should one measure instead if one wants to exploit the bound (\ref{bound})? After all, with the assumption of independent states the bound is correct. One could perform quantum tomography \cite{tomo}, i.e., estimate a quantum state by performing sufficiently many, informationally complete, different measurements on one's systems, and then apply the bound to the reconstructed density matrix. This would defeat the main purpose of the direct measurement, which is, indeed, to avoid tomography. 
Alternatively, one could perform measurements that explicitly check for the independence of two copies 1 and 2. For example, one may measure complete sets of observables $\{O^k_1,O^k_2\}$ on copies 1 and 2, and verify that
$\langle O^k_1\otimes O^{k'}_2\rangle=\langle O^k_1\rangle \langle O^{k'}_2\rangle$  for all pairs $k,k'$.
But then one can reconstruct the density matrix from these data, thus  again defeating the purpose of the direct measurement.

Let us insist then on performing {\em just} direct measurements of entanglement.  As pointed out in \cite{renner}, if one generates {\em many} copies whose state lives in a 
{\em symmetric} Hilbert space, then one can  {\em derive} bounds on the independence and identity of a small number of copies. The symmetry of the multi-copy state is enforced by randomly permuting the states and only then performing the appropriate measurements.
(In the limit of infinitely many copies, the precise statement concerning the form of the joint state assigned to all copies is known as the quantum de Finetti theorem \cite{finetti}.) 

Let us see why producing multiple copies of the counter-example state $|\psi\rangle$ given above and permuting pairs indeed will lead to the correct estimate of entanglement between Alice and Bob, namely zero,  for a sufficiently large number of such states.
Suppose, then, Alice and Bob have $M=N/2$ copies of $|\psi\rangle$ (thus  $N$ bipartite states have been generated).
They each apply the same randomly chosen permutation to their systems, and then perform a measurement of $V_i$ on the first pair of bipartite systems. The probability that that pair is
{\em not} in the joint state $|\psi\rangle$ is $(N-2)/(N-1)$. In this case the state is actually just the maximally mixed state (tracing over particles 1 in the state $|\psi\rangle$ leaves particles 2 in the maximally mixed state): in that state, the expectation value of $V_i$ is negative: $\langle V_i\rangle=4\times(1/4-3/4)\times1/4=-1/2$.
With probability $1/(N-1)$ the joint state of the two pairs is of the form $|\psi\rangle$, in which case the expectation value is, as before,  
$\langle V_i\rangle=4$. Thus, the expectation value of $V_i$ is the weighted average $\langle V_i\rangle=(5-N/2)/(N-1)$. The correct conclusion that there is no entanglement will be reached as soon as $N\geq 10$ in this special case. This illustrates the benefits of permutation symmetry.

Now consider the general case, where we do not have any information about the states of our pairs. We do not assume that each pair of bipartite systems is in the state $|\psi\rangle$, nor do we assume that they are all independently and identically distributed.
Instead, we will make use of Renner's theorem \cite{renner}, which makes a {\em qualified} assumption about the form of the states of multiple copies. That is, every statement will be accompanied by an upper bound on the error on the state assignment. Here is what one can do in the specific case one performs measurements of the operators $V_i$:

{\bf (i)} Generate a large number $N$ of (entangled) bipartite systems, $k=1\ldots N.$ These systems must exist at the same time for step (ii) to be possible.

{\bf (ii)} Perform the desired joint measurements $V_i$ on a smaller number $n:=N-K$ of {\em randomly} chosen pairs of systems $(k_1,k_2)$ with $k_1\neq k_2$. 

{\bf (iii)} One may now {\em tentatively} assign a state to the $n$ bipartite states
that is of the form $\rho_0^{\otimes n-r}\otimes \rho^r$ for some permutation of the $n$ states (we do not specify any particular $\rho_0$, nor any particular $\rho^r$, nor any particular permutation).
That is, $n-r$ systems are in states of the desired independent form (in fact, even more than that, they are identical), and $r$ copies are ``bad copies'' whose overall state does not factor and may contain correlations and entanglement between copies.

{\bf (iv)} The previous state assignment comes with an error, which depends on one's choice of $N$, $K$, and $r$. This error is bounded from above by \cite{renner}
\begin{equation}\label{error}
E=3K^4\exp(-K(r+1)/N).
\end{equation}
The error refers to the distance between the tentative state assigned to the $n$ systems and the ``actual'' state which we would assign if we would do full quantum state tomography.

This procedure then must be repeated sufficiently many times in order to obtain reliable statistics for determining
the two quantities $\langle V_i\rangle$ for $i=1,2$.

Suppose one measured an average value $V_m$ for one of the operators $V_i$ (in step (ii)).
The probability that one has picked a pair for which the density matrices do {\em not} factor, according to the state assignment made in (iii), is $P_{bb}=r(r-1)/n(n-1)$.
Indeed, there is no factorization only when {\em both} states are in the group of $r$ ``bad copies.''
For these ``bad copies'' we have to assume the {\em worst case}. Let us assume the bad copies are in the counter example state $|\psi\rangle$ if we have at most 10 bad copies, and in some unentangled state otherwise. For the bad copies we thus assume an expectation value of $V_i$ given by (see above)
\begin{eqnarray}
V_b(r)&=&(5-r/2)/(r-1)\, {\rm for}\, 2\leq r\leq 10\nonumber \\
V_b(r)&=& 0 \,\,{\rm for}\, r>10.
\end{eqnarray}
The remaining fraction $1-P_{bb}$ of ``good copies'' does satisfy a bound of the form (\ref{bound}).  If we denote by $C_0$ the concurrence of the state $\rho_0$, then we have the bound
\begin{equation}
C_0^2 \geq \min\left(\frac{V_m-P_{bb}V_b}{P_n},1\right):=C_{\min}
\end{equation}
where
$
P_n=(n-r)(n-r-1)/n(n-1)
$
is the probability to pick two good copies.

If we are interested in the {\em average} concurrence, $\bar{C}$, of all $n$ copies (after all, we do not know which are good copies and which are bad), we get the bound
\begin{equation}\label{conc}
\bar{C}\geq \frac{n-r}{n}C_{\min}
\end{equation}
The above procedure thus produces an estimate of a lower bound on the average concurrence, (\ref{conc}), and an upper bound $E$ on the error we make in our state assignment, (\ref{error}).
Let us now analyze how to pick reasonable values of $N,K,r$ (and after that we will optimize those choices). First of all, in order to decrease the error $E$ we have to discard a large number $K$ of bipartite systems. But we also have to be modest in our choice of the number of systems $n-r$ that we can assume are independent and identical. That is, we also must choose $r$ large.
On the other hand, in order for our estimate of the concurrence (\ref{conc}) to be reasonable, we cannot choose $r$ too large either.

Thus, let us choose $r\leq K\leq N$, such that in the limit of $N\rightarrow\infty$ we have both $r/N\rightarrow 0$ and $K/N\rightarrow 0$. In that case, the estimated concurrence obeys 
\begin{equation}
\bar{C}\rightarrow \sqrt{V_m} \,\,{\rm for} \,N\rightarrow\infty.
\end{equation}
Let us make the following somewhat arbitrary choice:
take $K=N^\beta$, with $\beta$ somewhere between 0.5 and 1. Similarly, choose $r=(N-K)^\alpha$ with $\alpha$ somewhere between 0.5 and 1.
Let us then vary the values of $\alpha$ and $\beta$ to see how the error behaves as a function of $N$, and how the estimate of concurrence behaves.
Examples are given in Figure 1.
\begin{figure}
  \includegraphics[width=9cm]{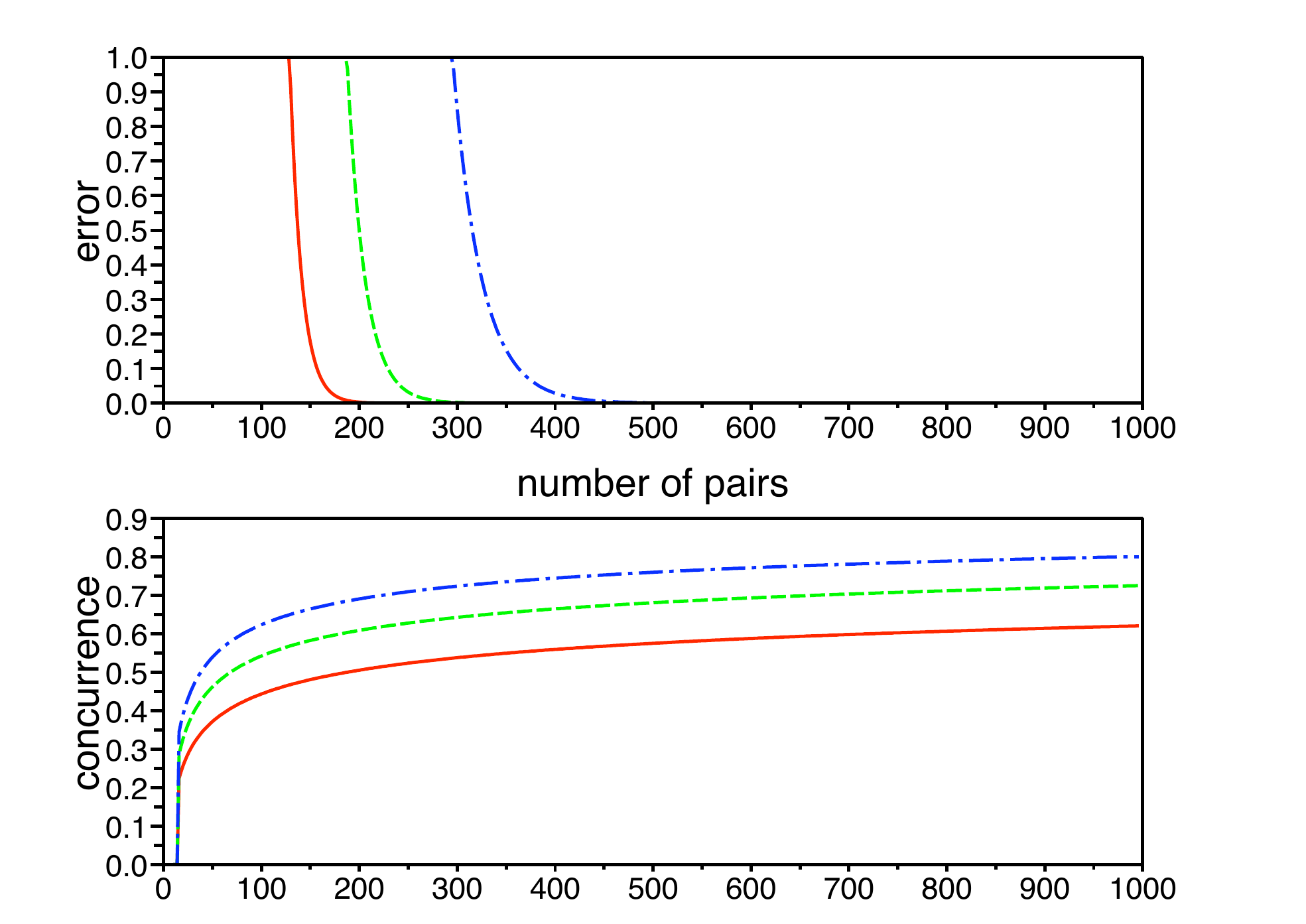}
  \caption{Upper bound $E$ on the error, given by Eq. (\ref{error}) (top figure) and estimated concurrence $\bar{C}$ (lower figure), as given by Eq. (\ref{conc}), as functions of the number of copies produced, $N$, for different choices  $r=(N-K)^\alpha$. Here, $K=N^\beta$, with $\beta=0.85$, and  $\alpha=0.75$ for the blue dash-dotted curves,  $\alpha=0.8$ for the green dashed curves, and $\alpha=0.85$ for the red solid curves.
It is assumed here that the measured value $V_m$ obeys $\sqrt{V_m}=0.8$. The estimated concurrence $\bar{C}$ reaches $\sqrt{V_m}$ asymptotically for $N\rightarrow\infty$. }
\end{figure}
We fixed $\beta=0.85$  and varied
$\alpha$. 
The larger $\alpha$ we choose, the more states we assume are ``bad,'' and the quicker the upper bound to the error in that statement decays to zero. On the other hand, the estimate of the concurrence approaches the correct value (here, $\sqrt{V_m}=0.8$) earlier for smaller values of $\alpha$. There is thus a compromise between a good estimate for the concurrence and a small state-assignment error.

For the largest  value of $\alpha$ plotted, $\alpha=0.85$, one needs about $N=200$ generated copies for the error to become sufficiently small, and yet have a reasonable lower bound on the concurrence (in this case, $\approx 0.5$). For even larger values of $\alpha$ (not plotted) the error tends to zero for smaller values of $N$, but the estimate of the concurrence will be smaller (reaching zero, eventually). 

Let us now fix a number $N$, insist on a certain maximum error $E$, and then find the best (highest) estimate of the concurrence, consistent with $N$ and $E$. 
Results for $N=100$ and $200$ are plotted in Figures 2--3.
These figures confirm that it is sufficient to generate $N=200$ pairs at the same time 
to produce a good estimate of the concurrence, whereas $N=100$ pairs is not quite sufficient. Obviously, this makes it a challenge to implement direct measurements in practice.
\begin{figure}
  \includegraphics[width=8cm]{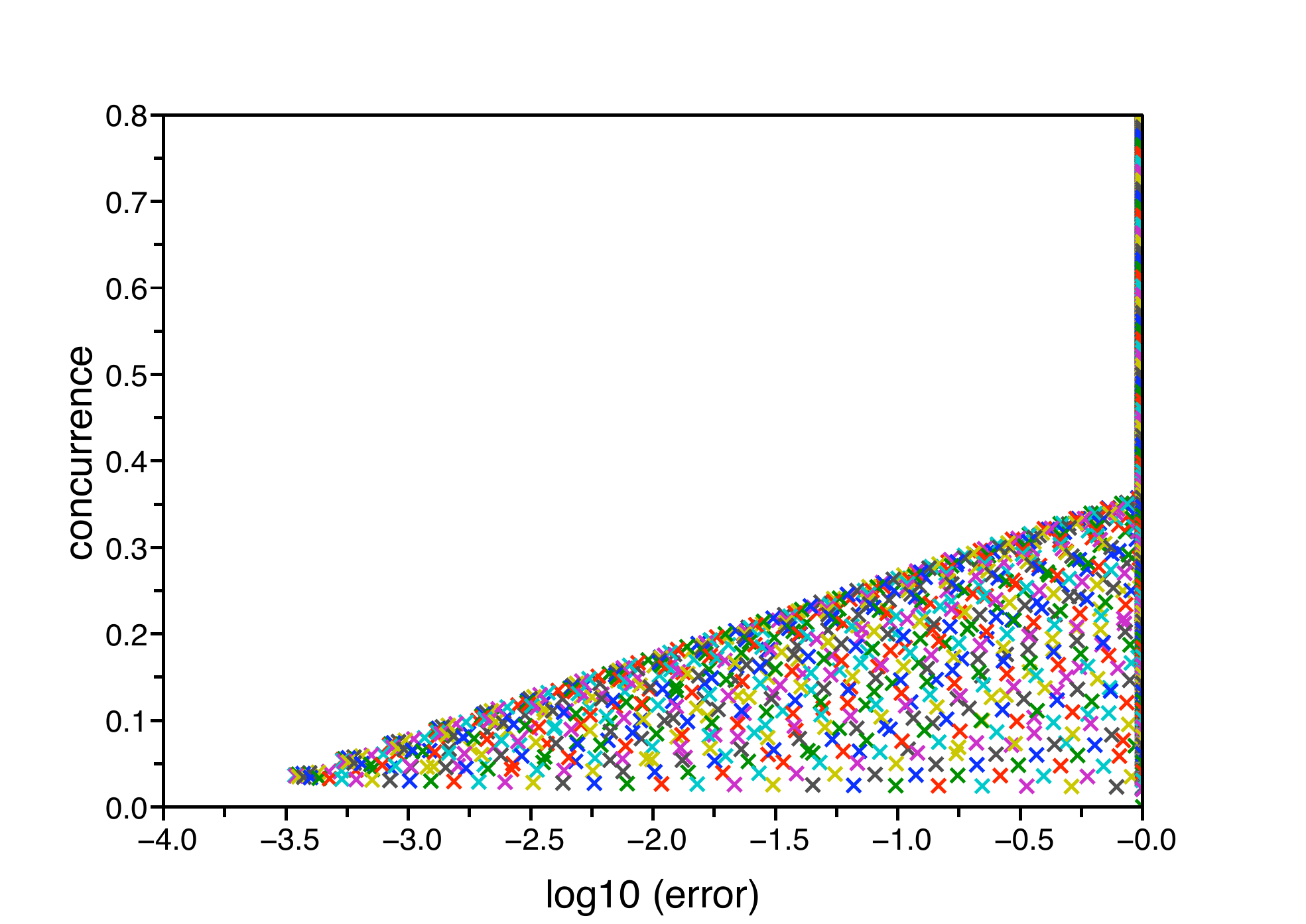}
  \caption{Scatter plot of the lower bound on the concurrence, $\bar{C}$ vs. the logarithm (base 10) of the upper bound $E$ on the error, for {\em all} possible values for $k$ and $r$, given a fixed value of $N=100$, and  assuming the measured value of $\sqrt{V_m}=0.8$. All points together give rise to a tradeoff curve between a lower bound to the concurrence and its reliability: the larger one chooses that lower bound, the less reliable it is. $N=100$ is not quite sufficient to approach the correct value of the concurrence with appreciable certainty.}
\end{figure}
\begin{figure}
  \includegraphics[width=8cm]{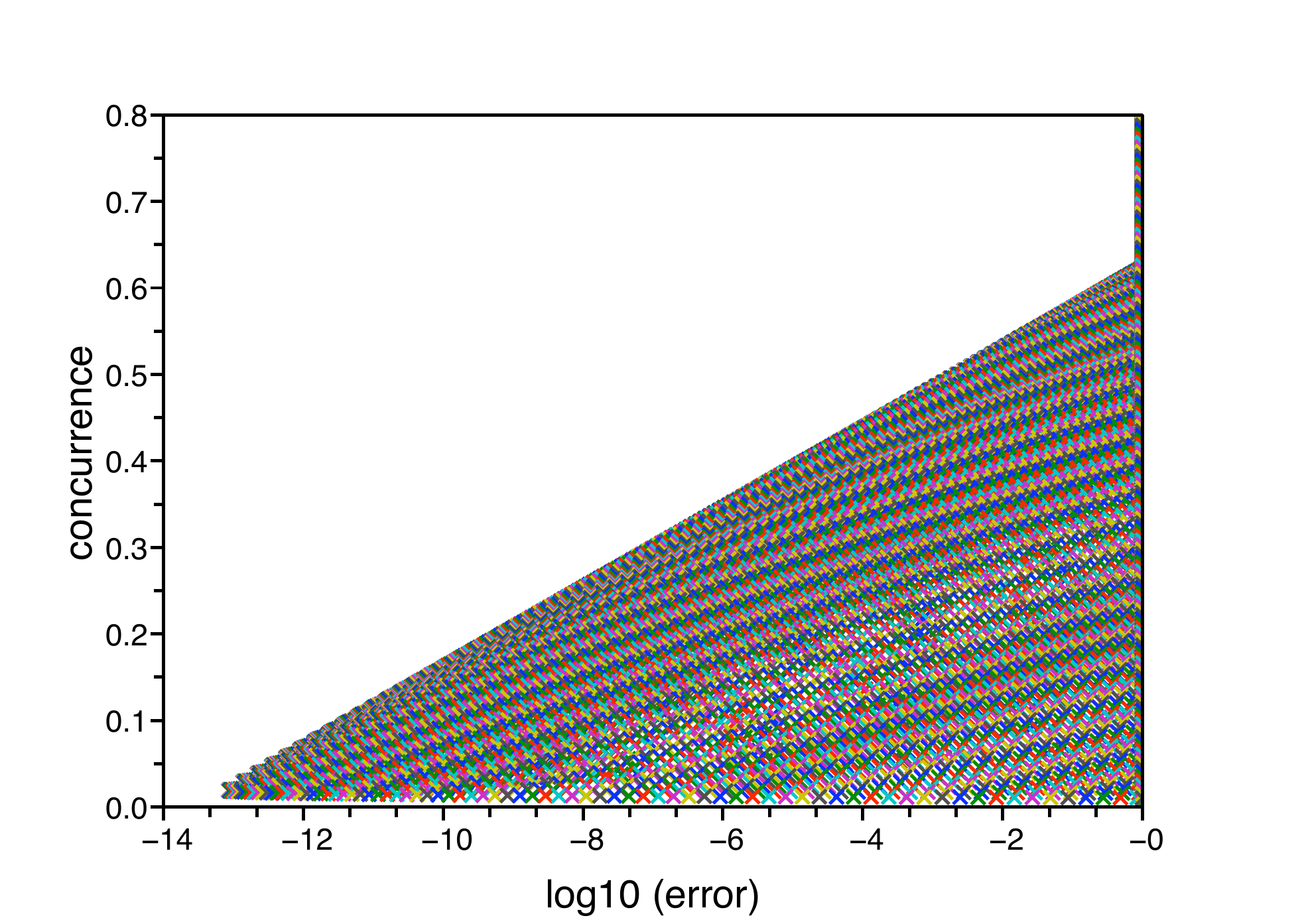}
  \caption{Same as Fig. 2, but for $N=200$. One can get a reasonable estimate of the concurrence with a small error in one's state assignment.}
\end{figure}
 
Let us finally compare the role permutation symmetry plays in direct measurements to those for entanglement witnesses and for tomography. For an entanglement witness, one measures just one observable on (an ensemble of) single copies. Thus, there is no need to do anything to enforce permutation symmetry. There is no reason to have multiple copies available {\em at the same time}. Thus, a measurement of an entanglement witness is much easier than a correctly implemented direct measurement where one needs to store a hundred copies or so of the systems to be tested. In order to perform quantum tomography 
one does have to perform different measurements, but they can all be done on single copies. In this case, in order to  enforce permutation symmetry on the Hilbert space of all copies, it would be necessary to perform the different measurements in random order on the single copies. But there is no need to have available the multiple copies {\em at the same time}. 

In conclusion, direct measurements of entanglement may provide necessary conditions for entanglement, but they are not sufficient. Turning direct measurements into actual entanglement verification tests (which are always sufficient) without additional (tomographic) measurements, requires substantial experimental effort: namely,  at least a hundred of instances of the entangled systems have to be generated and stored to be available for the direct measurements on randomly permuted pairs.

I thank Michael Raymer for useful discussions.

\end{document}